\documentclass[a4paper]{PoS}
\usepackage[utf8]{inputenc} %
\usepackage{amsmath} %
\usepackage{amsthm} %
\usepackage{mathtools} %
\usepackage{siunitx} %
\usepackage{booktabs} %
\usepackage{tikz} %
\usepackage{pgfplots} %
\pgfplotsset{compat=1.13} %
\usepackage[
bibencoding=utf8,
backend=biber,
style=phys,
maxnames=3,
sorting=none,
sortcites=true,
pageranges=false,
eprint=false,
articletitle=false,
biblabel=brackets,
]{biblatex} %

\let\originalleft\left
\let\originalright\right
\renewcommand{\left}{\mathopen{}\mathclose\bgroup\originalleft}
\renewcommand{\right}{\aftergroup\egroup\originalright}

\newcommand{\mc}[1]{\mathcal{#1}}
\newcommand{\diff}{\text{d}} %
\renewcommand{\Im}[0]{\text{Im}}

\renewcommand{\hat}[1]{\widehat{#1}}

\title{Quark mass dependence of $\gamma^{*}\pi\rightarrow\pi\pi$}

\ShortTitle{Quark mass dependence of $\gamma^{*}\pi\rightarrow\pi\pi$}

\author{\speaker{Malwin Niehus}\\
        Helmholtz-Institut für Strahlen und Kernphysik (Theorie) and\\
        Bethe Center for Theoretical Physics, Universität Bonn, D--53115 Bonn, Germany\\
        E-mail: \email{niehus@hiskp.uni-bonn.de}}

\author{Martin Hoferichter\\
        Institute for Nuclear Theory, University of Washington, Seattle, WA 98195-1550, USA \\
        E-mail: \email{mhofer@uw.edu}}

\author{Bastian Kubis\\
        Helmholtz-Institut für Strahlen und Kernphysik (Theorie) and\\
        Bethe Center for Theoretical Physics, Universität Bonn, D--53115 Bonn, Germany\\
        E-mail: \email{kubis@hiskp.uni-bonn.de}}

\abstract{Usually the simulation of scattering processes in lattice QCD is carried out at unphysically high values of the quark masses.
Hence, a method to extrapolate data obtained in lattice calculations to physical masses is needed to allow for comparison between theory and experiment.
To obtain a sound extrapolation, dispersion relations and chiral perturbation theory can be invoked.
While a simple combined approach known as the inverse amplitude method allows for a successful extrapolation of $\pi\pi\to\pi\pi$ data, a more complicated framework is needed for inelastic processes such as $\gamma^{*}\pi\to\pi\pi$.
By employing a well-established dispersive description, the extrapolation can be performed for $\gamma^{*}\pi\to\pi\pi$ both for on-shell as well as virtual photons, the decay $\gamma^*\rightarrow\pi\pi\pi$ is also within the range of applicability.
This particular process is interesting due to both its contribution to the anomalous magnetic moment of the muon and its connection to the axial anomaly.
}

\FullConference{The 9th International Workshop on Chiral Dynamics\\
    17--21 September 2018\\
    Durham, NC, USA}

\begin{document}

\section{Introduction}
Nowadays ab initio calculations of scattering processes using lattice quantum chromodynamics (QCD) are common, however, they are often carried out at unphysically high values of the quark masses.
Although the calculations are getting closer and closer to the physical point, i.e.\ physical quark masses, often extrapolation is needed to reach it (for a recent review of two-particle scattering on the lattice, see Ref.~\cite{Briceo2018}).
Moreover, to extract the interesting characteristics of the QCD spectrum from lattice QCD computations, parametrizations of the resulting data are required that allow for a continuation of the data to the complex plane, where the properties of resonances are encoded in pole positions and residues.
This remains true even with lattice data at the physical point.
Furthermore, there are ongoing efforts to make the calculation of three-particle scattering processes in lattice QCD feasible, which was for a long time restricted to processes with two particles in the initial and final states, respectively; see e.g.\ Refs.~\cite{Briceno:2018aml,Pang:2019dfe,Mai:2017bge}, for a review see Ref.~\cite{Hansen2019}.
Upcoming calculations of three-particle scattering are likely to increase the need for theoretically sound parametrizations.
Ideally, parametrizations tackling the aforementioned points should be based on fundamental theoretical principles and introduce as few model assumptions as possible.
Since one aims for a determination of pole positions of resonances, dispersion relations seem to be a natural candidate.
In combination with chiral perturbation theory (ChPT) they have already been applied to the process $\pi\pi\rightarrow\pi\pi$, both for extrapolation to the physical point and determination of resonance properties, see e.g.\ Ref.~\cite{Bolton2016}.
Here we propose a dispersive framework to describe the quark mass dependence of the process $\gamma^*\pi\rightarrow\pi\pi$.
This particular process is interesting for four reasons.
First, it is related to the anomalous magnetic moment of the muon, where a deviation between experiment and theory might provide a hint for physics beyond the Standard Model~\cite{Colangelo2014,Hoferichter2018}.
Second, the scattering amplitude at low energies can be predicted analytically using the axial anomaly in QCD~\cite{Adler1971}.
This prediction is currently tested only at the \SI{10}{\percent} level, to allow for a better check a dispersive description of $\gamma^*\pi\rightarrow\pi\pi$ was presented in Ref.~\cite{Hoferichter2012}, in fact, the framework presented here is an extension of this description.
Third, the dispersive approach provides access to the radiative couplings of the $\rho$ resonance, which is the dominating signature in the scattering process at hand in the energy region below \SI{1}{GeV}.
The benefit of a dispersive treatment regarding this aspect is that one can extract the radiative coupling directly from the residue at the pole position~\cite{Hoferichter2017}.
Last, a description of photon--pion scattering, being more complicated than $\pi\pi$ scattering in various aspects, provides the canonical next step towards more complicated processes, both from a lattice~\cite{Briceo2015,Briceno:2016kkp} and a dispersive point of view.
Currently, only two lattice calculations of $\gamma^*\pi\rightarrow\pi\pi$ exist, namely the ones in Refs.~\cite{Briceo2015,Briceno:2016kkp,Alexandrou2018}, carried out at $M_\pi\approx\SI{400}{MeV}$ and $M_\pi\approx\SI{320}{MeV}$, respectively.
The former one is probably beyond the breakdown-scale of the framework presented here, but upcoming calculations at lower pion masses will be in the range of applicability.
Since $\pi\pi$ scattering provides key input to the $\gamma^*\pi\rightarrow\pi\pi$ amplitude, in Sec.~\ref{sec:pipi} we will first discuss extrapolation of $\pi\pi$ scattering data.
Here we will also point out some issues in previous analyses of the data that make a re-analysis interesting in its own right.
Subsequently, in Sec.~\ref{sec:gammapi} the description of $\gamma^*\pi\rightarrow\pi\pi$ is presented, the framework is summarized in Sec.~\ref{sec:conclusion}.
The idea to use the quark mass dependence of $\pi\pi$ scattering as input for a more evolved dispersive framework dates back at least to Ref.~\cite{Guo2009}.
The flexibility of this approach is also underlined by the adaptation of the framework presented here to the decay of an $\omega(782)$ or $\phi(1020)$ into three pions in Ref.~\cite{Dax2018}.
Since ChPT relates the quark masses to the pion mass $M_\pi$ (henceforth we work in the isospin limit) and $M_\pi$ is the quantity that enters directly into the formulas, in the following the $M_\pi$-dependence is considered.

\section{Quark mass dependence of $\pi\pi\rightarrow\pi\pi$} \label{sec:pipi}
There exists a well-known framework to describe the $M_\pi$-dependence of $\pi\pi\rightarrow\pi\pi$ in the elastic regime (i.e.\ taking into account $\pi\pi$ intermediate states only), namely the inverse amplitude method (IAM)~\cite{Truong1991,Dobado1993,Dobado1997,GmezNicola2008}.
First, we briefly recapitulate this method focusing on the $I=J=1$ partial wave $t_1$, for it is the only one relevant in our description of $\gamma^*\pi\rightarrow\pi\pi$.
To start with, one expands $t_1$ in $SU(2)$ ChPT:
\begin{equation}
  t_1 = t_1^2 + t_1^4 + t_1^6 + \mc{O}\left(p^{8}\right).
\end{equation}
Here, $t_1^k$ is $\mc{O}(p^k)$ and $p$ represents the momenta and masses of the pions in units of the breakdown-scale of ChPT.
Furthermore, unitarity of the $S$-matrix implies
\begin{equation} \label{eq:pionpion_unitarity}
  \Im\left[t_1\left(s\right)\right] =
  \sigma\left(s\right)\left\vert t_1\left(s\right)\right\vert^2, \qquad
  \sigma\left(s\right) \coloneqq \sqrt{1-\frac{4M_\pi^2}{s}},
\end{equation}
for $s>4M_\pi^2$, $s$ being the square of the center of mass energy.
The ChPT expansion satisfies this unitarity relation only perturbatively, i.e.\ order by order in $p$.
In addition, it is not capable to describe resonances.
However, by combining the unitarity relation and the ChPT expansion with a dispersion relation for $t_1$ one obtains
\begin{equation} \label{eq:iam_nlo_amplitude}
  t_1
  \approx \frac{\left(t_1^2\right)^2}{t_1^2 - t_1^4}
  \eqqcolon t_1^\text{NLO}
\end{equation}
if one works to NLO in ChPT or
\begin{equation} \label{eq:iam_nnlo_amplitude}
  t_1
  \approx \frac{\left(t_1^2\right)^2}{t_1^2 - t_1^4 + (t_1^4)^2/t_1^2 - t_1^6}
  \eqqcolon t_1^\text{NNLO}
\end{equation}
if NNLO contributions are taken into account.
Eq.~\eqref{eq:pionpion_unitarity} plays a crucial role in the derivation of Eqs.~\eqref{eq:iam_nlo_amplitude} and~\eqref{eq:iam_nnlo_amplitude}, the derivation does not go through for processes where the unitarity relation takes on a different form, in particular it does not work for $\gamma^*\pi\rightarrow\pi\pi$.
It is worth noting that the IAM takes into account contributions from $\pi\pi$ intermediate states in the $s$-channel exactly, while such contributions in crossed channels are only considered up to the used order in ChPT.
Explicit expressions for $t_1^2$, $t_1^4$, and $t_1^6$ are obtained by projecting the scattering amplitude given in Refs.~\cite{Gasser1984,Bijnens1996} analytically onto the $p$-wave.
They depend on $M_\pi$, the pion decay constant $F$, and various low-energy constants (LECs).
While $t_1^2$ does not contain any additional LECs beside $M_\pi$ and $F$, in $t_1^4$ the linear combination $l_2^r - 2l_1^r$ appears and $t_1^6$ contains $l_1^r, l_2^r$, and $l_3^r$, as well as three linear combinations of NNLO LECs.
That is, $t_1^\text{NLO}$ contains one free parameter and $t_1^\text{NNLO}$ six free parameters.
Note that we work with the pion decay constant in the chiral limit.
This is beneficial for $F$ is independent of $M_\pi$.
A nice feature of the IAM is its analytic structure:
it has both a right- and a left-hand cut, although the latter
one is only given in its chirally expanded form.
Moreover, using the simple unitarity relation Eq.~\eqref{eq:pionpion_unitarity} one can analytically continue Eq.~\eqref{eq:iam_nlo_amplitude} as well as Eq.~\eqref{eq:iam_nnlo_amplitude} to the second Riemann sheet, where resonances appear as poles.
The strategy to extrapolate lattice data works as follows:
the LECs contained in the IAM amplitude are fixed by a fit to the data at unphysically high $M_\pi$, subsequently the pion mass in the IAM amplitude is set to its physical value.
In the fit one needs to face a difficulty:
instead of values of the partial wave $t_1$ or its phase $\delta$, the lattice calculations produce discrete energy values $E^\text{lat}$.
Energy levels $E$ are connected via a quantization condition of the form
\begin{equation} \label{eq:pipi_quantisation}
  \cot\left[\delta\left(E\right)\right] = \mc{Z}\left(E\right)
\end{equation}
to the scattering phase shift $\delta$ (cf.\ Ref.~\cite{Briceo2018} and references therein), where $\mc{Z}$ is a known expression depending on the lattice characteristics.
To perform the fit, one computes the phase $\delta$ using the IAM to either NLO or NNLO for some fixed values of the LECs.
The resulting phase is plugged into Eq.~\eqref{eq:pipi_quantisation} to compute the corresponding energies $E^\text{IAM}$, these energies are fit to the ones obtained on the lattice via varying the LECs to minimize
\begin{equation}
  \chi^2 \coloneqq \sum_{k,j}\left(E_k^\text{lat} - E_k^\text{IAM}\right)C^{-1}_{kj}\left(E_j^\text{lat} - E_j^\text{IAM}\right).
\end{equation}
Here the sum runs over the different energy levels obtained on the lattice and $C$ denotes the correlation matrix of those levels.
\begin{figure}[t]
  \begin{center}
    \input{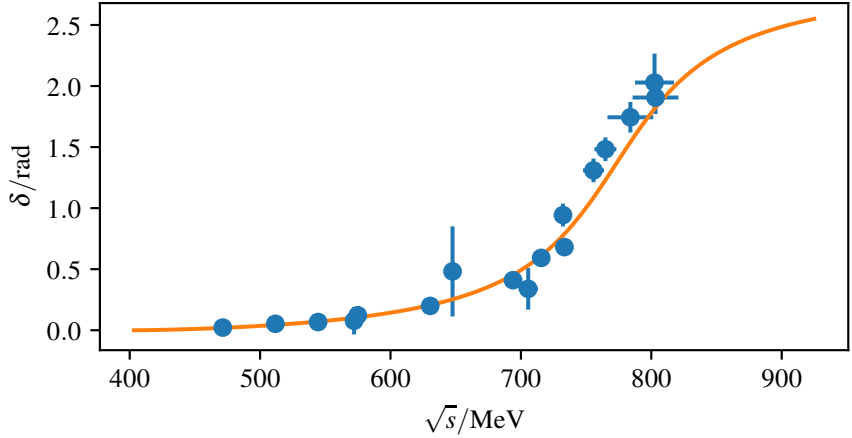}
  \end{center}
  \vspace*{-5mm}
  \caption{the result of the fit of the NLO IAM to the D200 ensemble from Ref.~\cite{Andersen2019} at $M_\pi\approx\SI{200}{MeV}$.} \label{fig:phase_fit}
\end{figure}
\begin{table}
  \begin{center}
    \begin{tabular}{ccccc}
      ensemble & $M_\pi/\si{MeV}$ & $\chi^2/\text{d.o.f.}$ & $p$-value & $48\pi^2(l_2^r-2l_1^r)$ \\
      \toprule
        N401 & 280 & 2.18 & \num{2.6e-3} & 6.3 \\
        N200 & 280 & 1.15 & 0.31 & 6.3 \\
        J303 & 260 & 0.84 & 0.65 & 5.8 \\
        C101 & 220 & 1.17 & 0.27 & 6.1 \\
        D101 & 220 & 2.48 & \num{3.5e-7} & 6.3 \\
        D200 & 200 & 0.69 & 0.81 & 5.8 \\
      \midrule
        20 & 390 & 1.95 & \num{2.9e-2} & 5.9 \\
        24 & 390 & 1.35 & 0.17 & 6.0 \\
        32 & 236 & 3.21 & \num{8.8e-7} & 5.9 \\
      \bottomrule
    \end{tabular}
    \caption{results of fits of the NLO IAM to lattice data where each ensemble is fit separately.
    The first six ensembles are from Ref.~\cite{Andersen2019} and the last three ones from Refs.~\cite{Dudek2013,Dudek2014,Wilson2015} (those are identified by their lattice length $L$).
    The pion masses are only approximate, the last column shows the obtained central values of the one LEC appearing at NLO.
    The pion decay constant is set to $F=\SI{86.67}{MeV}$~\cite{Tanabashi:2018oca,Aoki2017,Blum2016,Bazavov:2010hj,Borsanyi:2012zv,Durr:2013goa,Beane:2011zm}.} \label{table:nlo_fit}
  \end{center}
\end{table}
There are several lattice QCD calculations of the $\pi\pi$ $p$-wave available, up to now we analyzed data from Refs.~\cite{Dudek2013,Dudek2014,Wilson2015,Andersen2019,Knippschild2015}.
The results of fitting different data sets separately with the NLO IAM are summarized in Table~\ref{table:nlo_fit}, an example fit is shown in Fig.~\ref{fig:phase_fit}.
Although the $p$-value for some of the ensembles is acceptable, for others it is horrific, moreover, it is in general not true that lighter pion masses lead to a better fit.
However, note that the fit closest to the physical point is acceptable and yields a LEC compatible with the one that one obtains by fitting the NLO IAM to the experimentally observed phase.
Nevertheless, the NLO IAM clearly does not describe the lattice data sets satisfactorily.
This is not entirely unexpected, for the flexibility of the NLO IAM with only one free LEC is also insufficient to describe the experimental data perfectly, there is always a trade-off between a good agreement of the obtained mass or width of the $\rho$ with the observed values.
To improve the fit, in Ref.~\cite{Bolton2016} $F$ was replaced by its physical value, $F_\pi$, thereby introducing an additional LEC, $l_4^r$, to NLO.
While this yields a more satisfactory $\chi^2$, the resulting value of $l_4^r$ deviates significantly from its literature value, because this LEC is supposed to describe different physics from the one encoded in the $s$-dependence of the $\pi\pi$ phase shift.
Hence one is forced to use the NNLO IAM for a consistent fit.
To fix all six parameters, it is necessary to control both the $s$- and the $M_\pi$-dependence, that is ensembles with different pion masses need to be fit simultaneously.
Fortunately, recently published~\cite{Andersen2019} and upcoming~\cite{Knippschild2015} lattice calculations cover several different pion masses, so we hope to settle this matter soon.
It should be noted that there are past attempts to fit the NNLO IAM to lattice data~\cite{Pelez2010,Nebreda2011}, however, the lattice calculations used at that time are by now outdated, among other things because the $\rho$ was treated as stable.
Furthermore, Refs.~\cite{Hu:2016shf,Hu:2017wli} ascribe a significant impact on the $\rho$ characteristics to the strange quark, it remains to be seen whether the $SU(2)$ IAM is sufficient to describe the $N_f = 2+1$ lattice data used here.

\section{Quark mass dependence of $\gamma^*\pi\rightarrow\pi\pi$} \label{sec:gammapi}
\begin{figure}
  \begin{center}
    \input{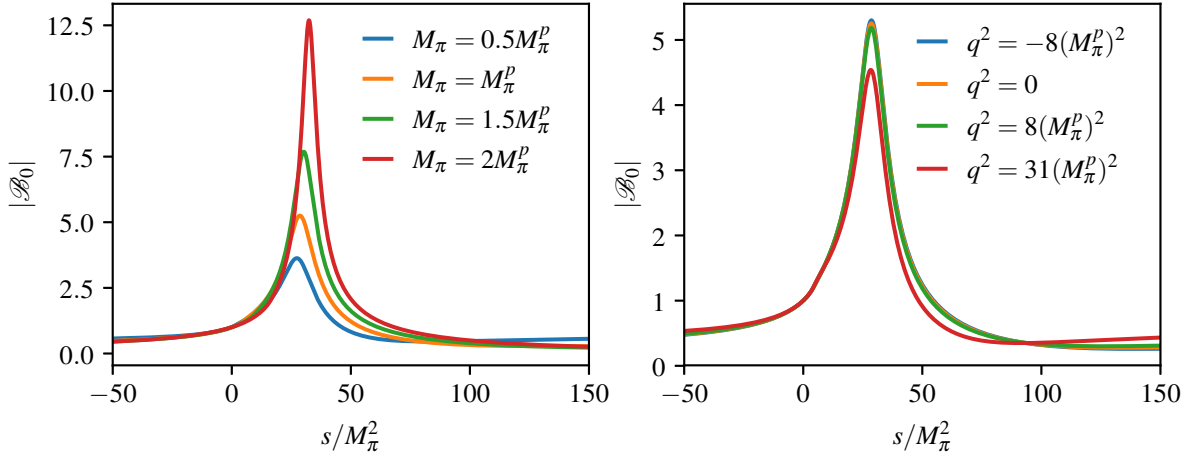}
  \end{center}
  \vspace*{-5mm}
  \caption{the absolute value of the basis function $\mc{B}_0$ along the real axis; the physical pion mass is $M_\pi^p = \SI{139.57}{MeV}$~\cite{Tanabashi:2018oca}.
  Left: different pion masses and $q^2=0$.
  Right: physical pion mass and different virtualities corresponding to spacelike, real, and timelike photons.} \label{fig:basis_functions}
\end{figure}
The scattering amplitude $\mc{M}$ for the process $\gamma^*\left(q\right)\pi^-\left(p_1\right)\rightarrow\pi^-\left(p_2\right)\pi^0\left(p_0\right)$ decomposes into a complex-valued function $\mc{F}$ and a prefactor dictated by parity according to
\begin{equation}
  \mc{M}\left(s,t,u; q^2\right)
  = i\epsilon_{\mu\nu\alpha\beta}\epsilon^\mu\left(q\right) p_1^\nu p_2^\alpha p_0^\beta \mc{F}\left(s,t,u; q^2\right),
\end{equation}
with $\epsilon(q)$ the polarization vector of the photon with 4-momentum $q$ (either on-shell or virtual) and the Mandelstam variables $s \coloneqq (q+p_1)^2$, $t \coloneqq (p_1-p_2)^2$, and $u \coloneqq (p_1 - p_0)^2$ constrained by $s + t + u = 3M_\pi^2 + q^2$.
As a consequence of isospin invariance as well as crossing symmetry the function $\mc{F}$ is completely symmetric in its arguments.
Its expansion into partial waves $f_J$ reads~\cite{Jacob1959}
\begin{equation} \label{eq:photon_pion_partial_waves}
  \mc{F}\left(s,t,u; q^2\right)
  = \sum_{J=0}^{\infty}f_{2J+1}\left(s,q^2\right)P_{2J+1}^\prime\left(z\right), \qquad z \coloneqq \cos\left(\theta\right).
\end{equation}
Here $\theta\coloneqq \angle(\vec{p}_1,\vec{p}_2)$ and $P_J^\prime$ are the derivatives of the Legendre polynomials.
Due to $G$-parity and Bose symmetry only odd partial waves contribute.
If we only consider $\pi\pi$ intermediate states, unitarity of the $S$-matrix enforces the lowest partial-wave, the $p$-wave, to obey~\cite{Hoferichter2012,Niecknig2012}
\begin{equation} \label{eq:photon_pion_unitarity}
  \Im\left[f_1\left(s,q^2\right)\right]
  = \sigma\left(s\right) f_1\left(s,q^2\right) t_1\left(s\right)^*.
\end{equation}
This unitarity relation is valid for $s>4M_\pi^2$ and relates the $p$-wave of $\gamma^*\pi\rightarrow\pi\pi$ with the $t_1$ wave of $\pi\pi\rightarrow\pi\pi$.
Since the left-hand side of Eq.~\eqref{eq:photon_pion_unitarity} is real, as a consequence one obtains Watson's theorem~\cite{Watson1954}: the phase of $f_1$ is the same as the phase $\delta$ of $t_1$.
It is this relation that makes it necessary to understand the $M_\pi$-dependence of pion--pion scattering before one is able to pin down the $M_\pi$-dependence of photon--pion scattering.
Henceforth we use the IAM to provide expressions for $t_1$ and $\delta$.
To obtain a theoretically sound representation of $\mc{F}$ that is capable to describe the $M_\pi$-dependen\-ce, it is expedient to study the analytic structure of $\mc{F}$.
To that end, first one restricts the photon 4-momentum to obey $q^2 < (3M_\pi)^2$.
That is, one does not allow the photon to decay into three pions.
This simplifies the analytic structure of $\mc{F}$ significantly, the case $q^2>(3M_\pi)^2$ is treated later using analytic continuation in $q^2$.
Since we allow for $\pi\pi$ intermediate states only, $\mc{F}$ has branch points in $s$, $t$, and $u$ at the two-pion threshold, $4M_\pi^2$.
One attaches branch cuts to these points that conventionally extend along the real axis to infinity.
Taking into account this analytic structure, one writes down a so-called fixed-$t$ dispersion relation for $\mc{F}$ that expresses $\mc{F}$ via integrals along its branch cuts, the integrands contain the imaginary part of $\mc{F}$.
To proceed further, one uses Eq.~\eqref{eq:photon_pion_partial_waves} to relate this imaginary part to the one of the $p$-wave according to
\begin{equation}
  \Im\left[\mc{F}\left(s,t,u; q^2\right)\right]
  = \Im\left[f_1\left(s,q^2\right)\right], \qquad s > 4M_\pi^2,
\end{equation}
i.e.\ one ignores the imaginary parts of the higher partial waves.
Thus $\mc{F}$ is related via Eq.~\eqref{eq:photon_pion_unitarity} to the $\pi\pi\rightarrow\pi\pi$ $p$-wave.
Expressing the imaginary part of $\mc{F}$ in the fixed-$t$ dispersion relation in this way and employing the symmetry of $\mc{F}$ one arrives at the reconstruction theorem~\cite{Hoferichter2012,Niecknig2012}:
\begin{equation} \label{eq:reconstruction_theorem}
      \mc{F}\left(s,t,u; q^2\right)
      = \mc{B}\left(s,q^2\right) + \mc{B}\left(t,q^2\right) + \mc{B}\left(u,q^2\right),
\end{equation}
where the function $\mc{B}$ decomposes as:
\begin{equation} \label{eq:sum_of_basis}
  \mc{B}\left(s,q^2,M_\pi\right)
  = \sum_{k=0}^m c_k\left(q^2,M_\pi\right)\mc{B}_k\left(s,q^2,M_\pi\right),
\end{equation}
with $c_k$ a-priori unknown expressions independent of $s$ and $\mc{B}_k$ the basis functions that describe the $s$-dependence and read
\begin{align} \label{eq:basis_functions}
  \begin{split}
    \mc{B}_k\left(s,q^2,M_\pi\right)
    &= \Omega\left(s,M_\pi\right)\left[
      s^k
      + \frac{s^m}{\pi}\int\limits_{4M_\pi^2}^\infty\frac{\hat{\mc{B}}_k\left(x,q^2,M_\pi\right)\sin\left[\delta\left(x,M_\pi\right)\right]}{\left\vert\Omega\left(x,M_\pi\right)\right\vert\left(x-s\right)x^m}\text{d}x \right], \\
    \hat{\mc{B}}_k\left(s,q^2,M_\pi\right)
    &= \frac{3}{2}\int\limits_{-1}^1\left(1-z^2\right)\mc{B}_k\left(t\left(s,z,q^2,M_\pi\right),q^2,M_\pi\right)\text{d}z.
  \end{split}
\end{align}
Here
\begin{equation}
  \Omega\left(s,M_\pi\right)
  = \exp\left[\frac{s}{\pi}\int\limits_{4M_\pi^2}^\infty\frac{\delta\left(x,M_\pi\right)}{x\left(x-s\right)}\diff x\right]
\end{equation}
is the Omnès function for the $\pi\pi\rightarrow\pi\pi$ $p$-wave scattering phase $\delta$, $m$ the number of subtractions, the Mandelstam variable $t$ is expressed in terms of the other kinematic quantities, and all hitherto suppressed pion-mass dependencies are shown explicitly.
To understand Eq.~\eqref{eq:basis_functions} better, note that the Omnès function corresponds to a summation of $\pi\pi$ rescattering in the $s$-channel, while the integral containing the function $\hat{\mc{B}}_k$ describes the $\pi\pi$-rescattering effects from the crossed channels.
The approximations that went into Eq.~\eqref{eq:basis_functions}, namely the inclusion of $\pi\pi$ intermediate states only and the focus on the $p$-wave, can be justified by noting that higher-energetic contributions are suppressed by the integration kernel and that at low energies the $p$-wave is dominant due to the $\rho$ resonance.
One appealing feature of the decomposition in Eq.~\eqref{eq:sum_of_basis} is that one can compute the basis functions as soon as the phase $\delta$ is known, that is without any input for $\gamma^*\pi\rightarrow\pi\pi$.
Depending on the value of $q^2$, the numerical solution of Eq.~\eqref{eq:basis_functions} becomes challenging:
as soon as $q^2 > M_\pi^2$, Mandelstam $t$ takes on complex values, this renders a naive numerical computation inefficient.
Moreover, if $q^2$ exceeds the three-pion threshold, the $z$ integration in Eq.~\eqref{eq:basis_functions} interferes with the branch cut of $\mc{B}_k$.
Although at the moment lattice calculations in the decay region do not exist, there are ongoing efforts to render them possible in the near future, for a recent review see Ref.~\cite{Hansen2019}.
Hence with an eye towards future applications it is desirable to allow for $q^2>(3M_\pi)^2$; cf.\ also Ref.~\cite{Dax2018}.
To that end Eq.~\eqref{eq:basis_functions} is analytically continued via the prescription $q^2\rightarrow q^2 + i\epsilon$.
The resulting equations are known as Khuri--Treiman equations~\cite{Khuri1960}.
The analytic continuation can be performed in different ways.
Usually, the contour in the $z$-integration in Eq.~\eqref{eq:basis_functions} is distorted.
However, the resulting integral shows several singularities that are challenging to deal with.
Therefore, recently in Ref.~\cite{Gasser2018} it was demonstrated how to continue the Khuri--Treiman equations for a different process, the decay $\eta\rightarrow \pi\pi\pi$, via distorting the $x$-integration contour while leaving the $z$-integration untouched.
The resulting equations were then translated into a matrix equation that was solved iteratively.
While this method does not lead to spurious singularities, it has the drawback that one needs to continue the phase $\delta$ into the complex plane, which is unphysical and therefore requires the use of ad-hoc parametrizations.
To circumvent this issue, we modify the method slightly by rewriting the integrand in Eq.~\eqref{eq:basis_functions} as follows~\cite{Hoferichter:2013ama}:
\begin{equation}
  \frac{\hat{\mc{B}}_k\left(x,q^2\right)\sin\left[\delta\left(x\right)\right]}{\left\vert\Omega\left(x\right)\right\vert\left(x-s\right)x^m}
  = \frac{\hat{\mc{B}}_k\left(x,q^2\right)\sin\left[\delta\left(x\right)\right]\sigma(x)e^{i\delta\left(x\right)}}{\left\vert\Omega\left(x\right)\right\vert\sigma(x)e^{i\delta\left(x\right)}\left(x-s\right)x^m}
  = \frac{\sigma\left(x\right)\hat{\mc{B}}_k\left(x,q^2\right)}{\left(x-s\right)x^m} \frac{t_1\left(x\right)}{\Omega\left(x\right)},
\end{equation}
where we used that for $x>4M_\pi^2$ the phase of the Omnès function equals $\delta$ and
\begin{equation}
  t_1\left(x\right)
  = \frac{\sin\left[\delta\left(x\right)\right]e^{i\delta\left(x\right)}}{\sigma\left(x\right)}, \qquad x > 4M_\pi^2.
\end{equation}
With the integrand rewritten in this way, we carry over the procedure described in Ref.~\cite{Gasser2018} to $\gamma^*\pi\rightarrow\pi\pi$ and $\gamma^*\rightarrow\pi\pi\pi$, but instead of needing to continue the phase $\delta$, we can now work directly with the partial wave $t_1$.
Moreover, the distortion of the $x$-contour is further simplified by the fact that $t_1(x)/\Omega(x)$ does not exhibit a right-hand cut.
On the downside, this improvement requires one to have an expression for $t_1$ available that is valid in the complex plane.
Here, this is no problem for we can use Eq.~\eqref{eq:iam_nlo_amplitude} or Eq.~\eqref{eq:iam_nnlo_amplitude} directly.
In other scenarios one could instead use amplitudes provided by Roy-like equations, cf.\ Refs.~\cite{GarcaMartn2011,Caprini2012} for $\pi\pi\rightarrow\pi\pi$.
We solve the resulting equations both iteratively and via matrix inversion and check that both solutions agree.
As a proof of principle, in Fig.~\ref{fig:basis_functions} we show basis functions with two subtractions computed for different virtualities and pion masses using the NLO IAM as input with the LEC set to \num{5.9}.
As a next step, as soon as a sound description of $\pi\pi$ via the NNLO IAM is available, the expressions $c_k$ in Eq.~\eqref{eq:sum_of_basis} need to be fixed via a fit to lattice data.
To that end, the $q^2$- and $M_\pi$-dependencies of $c_k$ are parametrized, reasonable parametrizations take into account the analytic structure in $q^2$, cf.\ Ref.~\cite{Hoferichter:2014vra} for an explicit expression.
Subsequently, the free parameters of the parametrizations are fixed in a fit.
The relation between continuum and lattice is more involved than Eq.~\eqref{eq:pipi_quantisation}, it contains both a matrix element associated with $\gamma\pi\rightarrow\pi\pi$ as well as the phase $\delta$~\cite{Briceno2015a}.
Both up to now available lattice calculations are carried out at $M_\pi > \SI{300}{MeV}$, but the NNLO IAM is known to break down for pion masses larger than $\SI{300}{MeV}$--$\SI{350}{MeV}$~\cite{Nebreda2011,Pelez2010}.
Upcoming simulations at lower pion masses promise reasonable fits.

\section{Conclusion} \label{sec:conclusion}
We have presented a framework that allows for the extrapolation of lattice QCD data for $\gamma^*\pi\rightarrow\pi\pi$ in the pion mass as well as the determination of resonance characteristics of the $\rho$ encoded in its pole position.
It is founded in dispersion relations and ChPT, both of which are well-established theoretical tools.
The fits to lattice QCD data provide challenges:
for a consistent description of $\pi\pi$ scattering via the IAM it seems to be necessary to include NNLO contributions, which leads to more free parameters.
Both to fix these parameters and to allow for a sound extrapolation of $\gamma^*\pi\rightarrow\pi\pi$ data, high-quality lattice calculations covering a range of pion masses are needed.
Fortunately, such calculations are by now carried out, more will be available soon.
The challenges one faces in a sound analysis of lattice results highlight the importance of both refined lattice calculations as well as a focus on the interplay of lattice, dispersion relations, and ChPT.

\acknowledgments
Finite-volume energy levels taken from Ref.~\cite{Dudek2013,Dudek2014,Wilson2015} were provided by the Hadron Spectrum Collaboration -- no endorsement on their part of the analysis presented in the current paper should be assumed.
In addition, we would like to thank ETMC, the authors of Ref.~\cite{Alexandrou2018}, as well as John Bulava for sharing lattice data with us.
Moreover, we would like to thank Raúl Brice\~no, Tobias Isken, Marcus Petschlies, Jacobo Ruiz de Elvira, Akaki Rusetsky, Martin Ueding, Markus Werner, and David Wilson for many helpful discussions.
Financial support by the Bonn--Cologne Graduate School of Physics and Astronomy (BCGS), by the DFG  and NSFC through funds provided to the Sino–German CRC 110 "Symmetries and the Emergence of Structure in QCD" (DFG Grant No.\ TRR110 and NSFC Grant No.\ 11621131001), and by the DOE (Grant No.\ DE-FG02-00ER41132) is gratefully acknowledged.

\printbibliography

\end{document}